# Broadband emitter localization and prefabrication two-photon interference from InAs/InP quantum dots monolithically grown on an InGaAsP/InP distributed Bragg reflector

Paweł Mrowiński[1,*], Jan Śmigiel[1,2], Mikołaj Badura[2], Beata Ściana [2], Wojciech Kijaszek[2], Emilia Zięba-Ostój[1], Paweł Holewa[1] and Marcin Syperek[1]

[1]*Department of Experimental Physics, Faculty of Fundamental Problems of Technology, Wrocław University of Science and Technology, Wybrzeże Wyspiańskiego 27, 50-370 Wrocław, Poland*

[2]*Faculty of Electronics, Photonics and Microsystems, Wrocław University of Science and Technology, Janiszewskiego 11/17, 50-372 Wrocław, Poland*

** corresponding author: pawel.mrowinski@pwr.edu.pl*

**Abstract**

We demonstrate a monolithic telecom C-band single-photon platform based on low-density InAs/InP quantum dots grown on an InGaAsP/InP distributed Bragg reflector, with simulated upward light-extraction efficiency exceeding 10% within the 40 nm spectral band. Enhanced light extraction is crucial for wide-field photoluminescence imaging, enabling approximately 3 nm statistical centroid-localization precision of QD emitters due to improved signal-to-noise ratio. The identified charged exciton emitting at 1549.5 nm exhibits single-photon emission purity up to 94.7%, validated by a raw peak-area integration under pulsed quasi-resonant excitation. Under these conditions, photons emitted sequentially from the unprocessed planar structure reveal a raw two-photon interference visibility of 51% of the photon wave packet within the 2 ns time window and 41% over a 4 ns integration window, all without Purcell enhancement typical for cavity-embedded emitters. Combining broadband extraction, high photon coherence, and emitter preselection on a native III-V wafer, this architecture paves the way for marker-based deterministic fabrication of high-quality monolithic nanophotonic devices.

## Introduction

A scalable solid-state single-photon generator for quantum photonics must deliver pure, indistinguishable single photons at a rate adequate for practical use. Self-assembled semiconductor quantum dots (QDs) enable triggered photon generation and are compatible with nanophotonic integration [1–3]. Those emitting in the telecom C-band also enable direct application in low-loss silica-fiber network infrastructure, for multinode quantum information processing without frequency conversion. However, for self-assembled QDs, scalable source fabrication is inherently sequential due to the spatial and spectral distribution of emitters. Identification and localization of a suitable emitter in an unpatterned structure are required before fabricating an aligned photonic environment (e.g., a cavity) to direct photons into a selected mode, enhance extraction efficiency, and, in resonant devices, accelerate radiative decay via the Purcell effect. Consequently, the starting-material platform must maintain high photon quality, provide detectable photon flux for emitter preselection, and remain compatible with deterministic post-localization processing.

One of the first challenges is achieving a low density of QDs while maintaining control over their optical properties during epitaxial growth using molecular beam epitaxy (MBE) or metal-organic vapor-phase epitaxy (MOVPE). Among various approaches, QD density and spatial distribution can be controlled using buried stressors enabling site-controlled growth [4,5], local-droplet etching followed by nanohole filling [6–8], droplet epitaxy [8,9] or growth on dielectric masks [10]. For strain-driven Stranski–Krastanov (SK) growth, optimization of the growth conditions or post-growth annealing is typically required to reduce QD density. Different strategies have been investigated for InP-based QDs grown by MBE [11,12] and by low-pressure MOVPE directly on InP [13] .

However, even when a low-density QD layer is achieved, the random spatial distribution of the emitters presents an additional challenge for their integration into photonic structures with high manufacturing yield. Consequently, intermediate steps are required for QD preselection and spatial localization with respect to lithographic markers [14,15] . Alternatively, far-field optical lithography [16] or cathodoluminescence-assisted in situ electron-beam lithography [17] can assist the process.

Beyond identifying and locating a suitable emitter, efficient photon extraction is essential both for reliable optical preselection and for obtaining sufficient photon rates in subsequent quantum-optical experiments. Efficient coupling into free-space or on-chip optical channels is also crucial for practical single-photon sources. To improve free-space outcoupling, several approaches have been demonstrated, including bottom mirrors such as distributed Bragg reflectors (DBRs) [18–21] or metallic reflectors [22], often combined with microlenses [21], Gaussian-shaped mesa structures [23,24] or nanowires [25]. Alternatively, resonant photonic structures such as micropillars [26–28], microdisks [29], circular Bragg gratings [30,31], Moiré cavities [32], or two-dimensional photonic-crystal slabs [33] can be employed. On-chip coupling approaches include direct QD integration into waveguides [34], heterogeneous integration [35,36] and cavity-assisted waveguide coupling [37]. Typically, resonant cavity-based systems require accurate spatial and spectral matching between the QD and the optical mode and often operate in a narrowband regime. In contrast, reflector-assisted or nonresonant extraction structures can provide broadband enhancement of photon collection without necessarily introducing substantial Purcell enhancement.

Several self-assembled epitaxial QD material systems have been investigated for direct C-band emission, notably InAs/InP [38], In(Ga)As/GaAs structures employing strain-reducing layers [39–42], or recently InAs/InAlGaAs [31]. The former have demonstrated single-photon emission with low multi-photon generation probability, two-photon interference, coupling to nanobeams, and deterministic device integration [11–13,43,44]. More recently, an InAs/InAlGaAs QD integrated into a circular Bragg grating cavity demonstrated a raw two-photon interference visibility exceeding 90% [31] . These advances highlight the performance achievable in integrated telecom-band single-photon sources, while raising the question of how much photon coherence is already available in the epitaxial structure before nanofabrication.

Hong-Ou–Mandel (HOM) interference can serve as a rigorous benchmark for assessing photon coherence in pre-fabrication platforms. In cavity-backed devices, Purcell-enhanced radiative decay can improve photon indistinguishability by reducing the effects of dephasing [45]. Measurements carried out in unprocessed planar structures without intentional Purcell enhancement instead evaluate intrinsic coherence before final device fabrication. Optically triggered InAs/InP QDs in mesas with a metallic back-reflector have demonstrated full-wavepacket photon indistinguishability of 35±3%, exploiting an optical transition with a relatively short lifetime of 340 ps [46]. By contrast, substantially higher HOM visibility values of 71.9±0.2% [47] and 91.7±0.2% [31] have been reported only for cavity-integrated C-band emitters with Purcell-shortened lifetimes. Therefore, the achievable coherence in planar structures remains an open question.

This work presents a monolithic telecom-C platform utilising low-density InAs/InP QDs grown on top of a 20.5-pair InGaAsP/InP DBR. The QD position and upper InP spacer are engineered for broadband upward photon extraction without intentional Purcell enhancement. The calculated photon extraction efficiency is approximately 13%. This enhancement enables wide-field PL imaging of individual C-band emitters with a statistical centroid-localization precision of approximately 3 nm for the investigated QD. HOM interference is employed to benchmark the photon quality retained by the unprocessed planar structure. A charged-exciton transition at 1549.5 nm exhibits a raw $g^{(2)}(0)$ of 0.059±0.002 under pulsed quasi-resonant excitation. Consecutive photons display a raw two-photons interference visibility of approximately 41% over the full wavepacket, increasing to 51% within a 2 ns integration window. To the best of our knowledge, this result is among the highest raw full-wavepacket HOM visibilities reported for an unprocessed planar InAs/InP C-band QD platform without cavity-induced radiative rate enhancement. These results, which combine broadband extraction, emitter localization in the planar structure, and comparatively high photon coherence retained by the unprocessed platform, establish our monolithic QD-on-DBR system as a promising starting platform for deterministic monolithic telecom C-band nanophotonic devices.

## Experimental section

To provide elevated upward photon flux for QD preselection while retaining a native III-V platform, we developed a monolithic MOVPE-grown structure with low-density InAs/InP QDs deposited on a 20.5-pair InGaAsP/InP DBR (**Figure 1(a)**). Prior to DBR growth, a series of InGaAsP and InP calibration layers were prepared to determine the growth rates and the structural and optical parameters required for the DBR design. The InGaAsP growth conditions were iteratively adjusted using feedback from high-resolution X-ray diffraction [48] and ellipsometry to maximize the refractive index contrast (Δn) while maintaining near-lattice matching to InP, as well as keeping the InGaAsP bandgap above 0.9 eV to prevent overlap with the QD photoluminescence. The analysis yielded an InGaAsP composition of $In_{0.7}Ga_{0.3}As_{0.64}P_{0.36}$, and the refractive indices at 1550 nm of $n_{InGaAsP}$ = 3.432 and $n_{InP}$ = 3.169. The resulting index contrast, Δn = 0.263, was used together with the calibration growth rates to define nominal quarter-wave thicknesses of 112.9 nm for InGaAsP and 122.3 nm for InP. These parameters were used as inputs to 3D FDTD simulations to design the complete structure, determine the optimized DBR-to-QD layer distance and the upper InP spacer thicknesses, and to maximize emission collection efficiency near 1550 nm while maintaining broadband reflectivity in the range of 1.5-1.6 μm (Supplementary Information, Sec A2). The DBR and semiconductor/air interface form a low-finesse planar cavity with a calculated near-unity Purcell factor (Supp. Inf., Sec. A2, Fig. S1c).

Based on this design, the complete structure was monolithically fabricated on an InP(100) substrate. The final DBR is roughly 4.8 μm thick and terminates with an InGaAsP layer, followed by an InP spacer (nominal 181 nm), an optically active InAs QD layer grown in Stranski-Krastanov mode, and an InP cap (nominal 247 nm). The complete structure ends with a second, uncapped, optically inactive InAs QD layer, deposited to assess QD density and morphology by atomic force microscopy (AFM). The whole sample layout is schematically demonstrated in **Fig. 1(a)** together with a cross-sectional scanning electron microscopy (SEM) image of the fabricated structure. The SEM image shows a complete multilayer stack, confirming uniform InGaAsP/InP DBR interfaces. SEM image analysis (Supp. Inf., Sec. A7, Fig. S7) yielded layer thicknesses of 114.3±1.6 nm and 121.7±1.6 nm for InGaAsP and InP, respectively. AFM analysis (**Figure 1(b)**) reveals low-density InAs QDs of $1.3\times10^{8}$ $cm^{-2}$. The surface QDs exhibit a mean height of 18.2 nm with a standard deviation of 1.0 nm and characteristic lateral dimensions of approximately 85±5 nm along [110] and [1-10] and 72±5 nm along [100] and [010] directions, revealing a high-symmetry square-based nanostructure shape, challenging to achieve in the SK growth mode. Although the QD geometrical shape is close to C4V in-plane symmetry, the general symmetry of the zinc-blende crystal lattice of such a QD is C2V, while at the atomistic level, the [110] and [1-10] directions are not equivalent, leading to non-zero fine structure splitting of the excitonic state [49]. Moreover, atomic decomposition at interfaces can occur during growth of the InP capping layer, further reducing the system's symmetry.

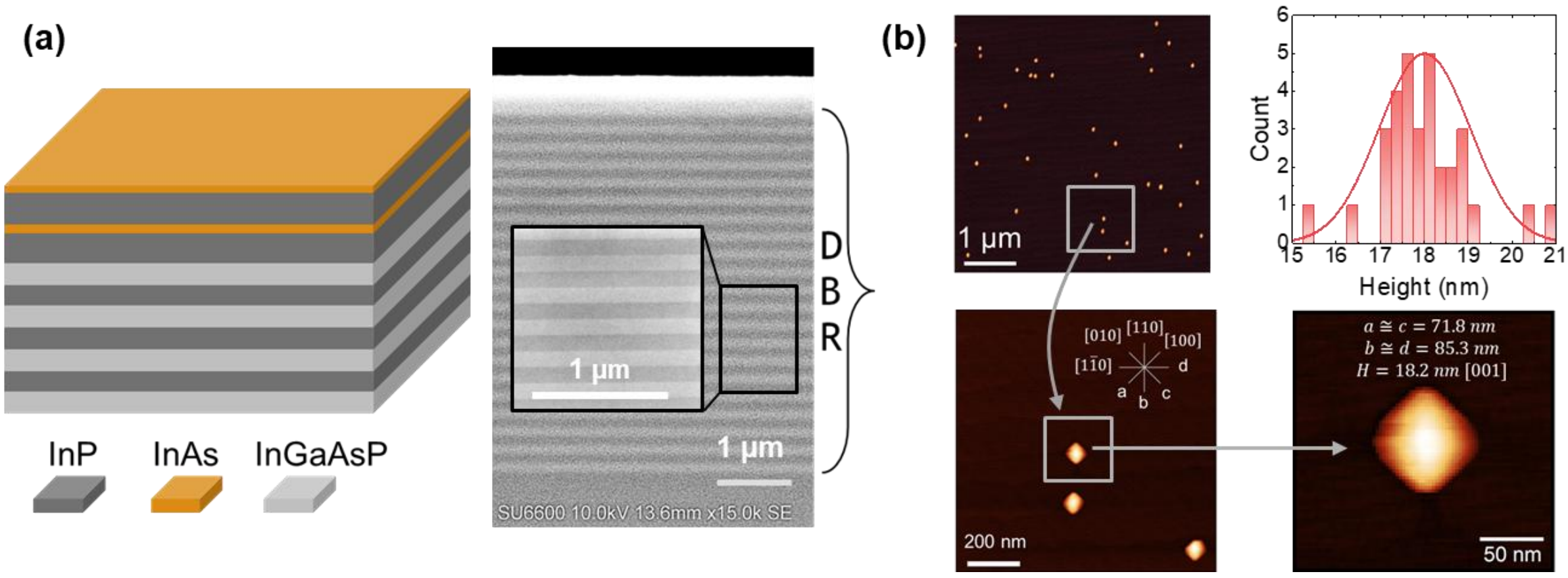

Figure 1. (a) Schematic layout of the planar sample composed of InGaAsP/InP DBR below the InAs/InP QD layer and the corresponding cross-sectional SEM image. b) AFM images of the uncapped QD layer and the distribution of QD heights with the fitting function (solid line).

**FDTD simulations and optical spectroscopy**

The optical response of the fabricated complete structure with a DBR was first evaluated by room-temperature reflectivity measurement using a silver mirror as a reference. The measured relative reflectivity reaches about 99% at 1533 nm, and the stopband is centered near 1542 nm with a width of roughly 70 nm (**Figure 2(a)**). The measured reflectivity spectrum is closely reproduced by the simulated multilayer response, taking into account refractive indices obtained from the ellipsometry analysis at room temperature (**Figure 2(b)**). A spectral offset of approximately 15 nm remains between the simulated and experimental responses. The discrepancy may reflect deviations of the fabricated layer thicknesses from the nominal design and uncertainties in the optical constants used in the simulations. To evaluate the influence of the DBR reflectivity on QD emission, an in-plane dipole was placed at the QD layer position in the 3D FDTD model. The DBR redirects emission propagating toward the substrate into the upper half-space, producing a calculated maximum upward photon extraction efficiency (PEE) of roughly 13% at 1573 nm, around four times the value obtained for an equivalent dipole in bulk-like InP (**Figure 2(d)**, see also Supp. Inf., Sec. A2, Fig. S1b). At high excitation power, the structure shows broadband photoluminescence (PL), whereas reducing the excitation power by two orders of magnitude resolves isolated excitonic transitions (**Figure 2(c)**). The single QD emitter spectral position (1549 nm) corresponds to a simulated upward photon extraction efficiency of roughly 10%. The DBR therefore operates as a broadband light extractor, providing the photon flux required for subsequent preselection of individual C-band emitters. The corresponding visualization of the upward directed field is demonstrated in **Fig. 2e)**.

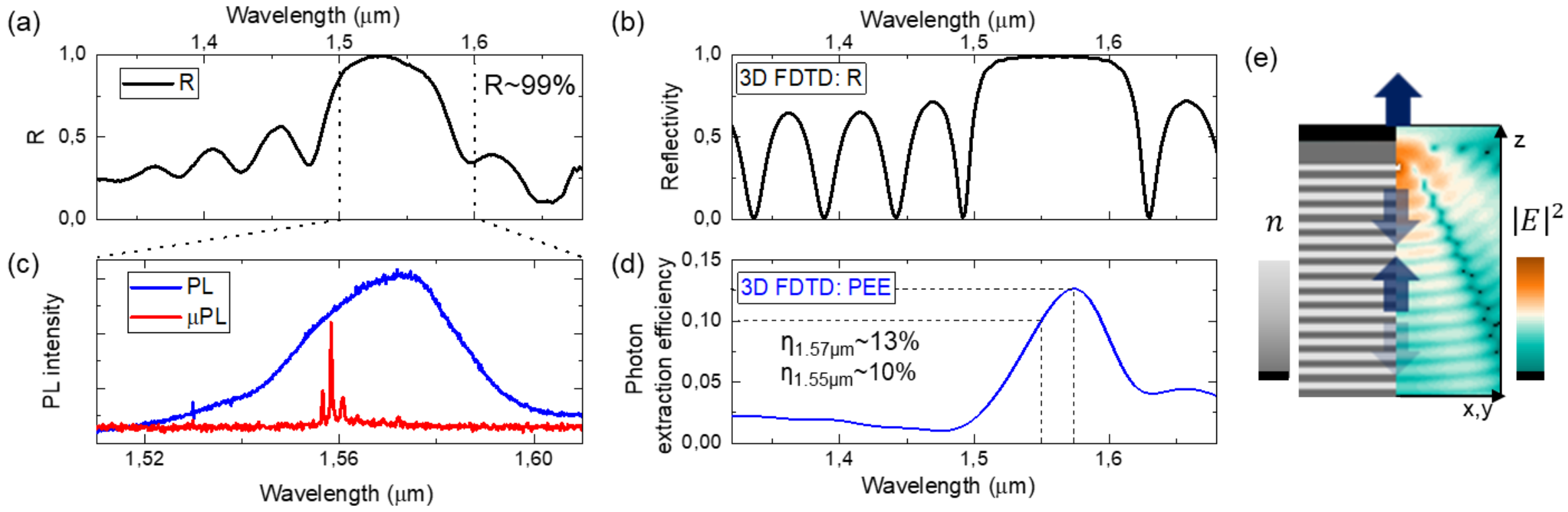

Figure. 2 Comparison of experimental (a,c) and simulated (b,d) results obtained for the low-density QD structure on a DBR. (a) Reflectivity (R) spectrum at room temperature. (c) PL from the QD ensemble and an example of the μPL spectrum from a single QD at T= 5 K. (b) Calculated reflectivity. (d) Photon extraction efficiency (PEE) with the embedded dipole source. (e) Simulated electric-field distribution across the sample cross-section influenced by the bottom mirror.

**Photoluminescence imaging**

Once the broadband photon extraction window is established, it is used to preselect individual C-band-emitting QDs. The preselection relies on wide-field PL imaging of the planar sample with spectral filtering. The sample was mounted in a closed-cycle cryostat at approximately 5 K and excited above the InP bandgap with an expanded laser beam focused on the back side of the objective to illuminate the entire objective field of view. PL was collected using an in-cryostat ×100 magnification objective (NA=0.7) and imaged onto a thermoelectrically cooled InGaAs camera (640×512 pixels, 15 μm each). Locating the objective inside the cryostat (**Figure 3(a)**) mitigates aberrations and blurring arising from light collection through the external cryostat window [15]. Moreover, the use of a high NA and short working distance objective elevates excitation power density and increases collection efficiency within the roughly 88 deg. solid angle (according to Snell's law applied in standard ray optics simulations, one can show that a beam angled 26 deg. to the normal direction refract to 88 deg. while 37.6 deg experience total internal reflection), both of which improve the signal-to-noise ratio (For more technical details see Supp. Inf., Sec. A3).

Emitter preselection was performed in two stages. A long-pass filter (λ > 1500 nm) was first used to identify a sample region with a limited number of optically active QDs, ensuring that PL background contribution from other QDs was negligible. The spectral window was then narrowed using approximately 12 nm-wide band-pass filters centered between 1550 and 1610 nm. **Figure 3(b)** presents a representative 80×64 μm$^2$ field acquired with a 1550/12 nm filter, in which spatially isolated QDs appear as diffraction-limited bright spots. Three emitters are revealed, whereas >100 spots are observed with the 1500 nm long-pass filter. This difference reflects the combined spectral selection imposed by the band-pass filter and the wavelength-dependent photon extraction provided by the DBR.

For quantitative localization, a 50×50-pixels region containing one isolated emitter was analyzed without image smoothing (camera-specific settings: 800 ms integration time, high conversion gain, fixed pattern noise correction, correlated double sampling for thermal noise reduction, see also Supp. Inf., Sec. A4). The intensity spot shows a weak Airy pattern around the central peak, which confirms that the wavefront phase coherence is mostly intact. The intensity distribution was fitted to a 2D Gaussian distribution, yielding a nearly circular point-spread function with a full width at half-maximum of approximately 9.9±0.1 pixels along x and 9.6±0.1 along y, corresponding to ~1.25 μm, being only ~1.2 times larger than the theoretical diffraction limit for our imaging system. The fitted centroid uncertainty is approximately 0.02 pixels. Since the calibration provides 125 nm/pixel, the uncertainty corresponds to roughly 2.5 nm. We conservatively quote a statistical localization precision of approximately 3 nm after accounting for the precision with which the fit uncertainty and pixel calibration are determined (see also Supp. Inf., Sec. A4, Fig. S3). The quoted value represents the statistical precision with which the center of the selected PL spot is determined. We underline here, that it is not yet the absolute QD-to-device placement accuracy. A complete deterministic fabrication workflow would additionally include localization of lithographic markers, correction of field-dependent optical distortions, coordinate transformation, and electron-beam lithography alignment. Nevertheless, the result demonstrates that the photon flux available from the unprocessed QD–DBR structure is sufficient for nanometer-scale emitter localization. The overall estimated offset accuracy evaluation for integration of subsequent nanophotonic structures would be $\pm$60 nm (see the proposed evaluation Supp. Inf., Sec. A4).

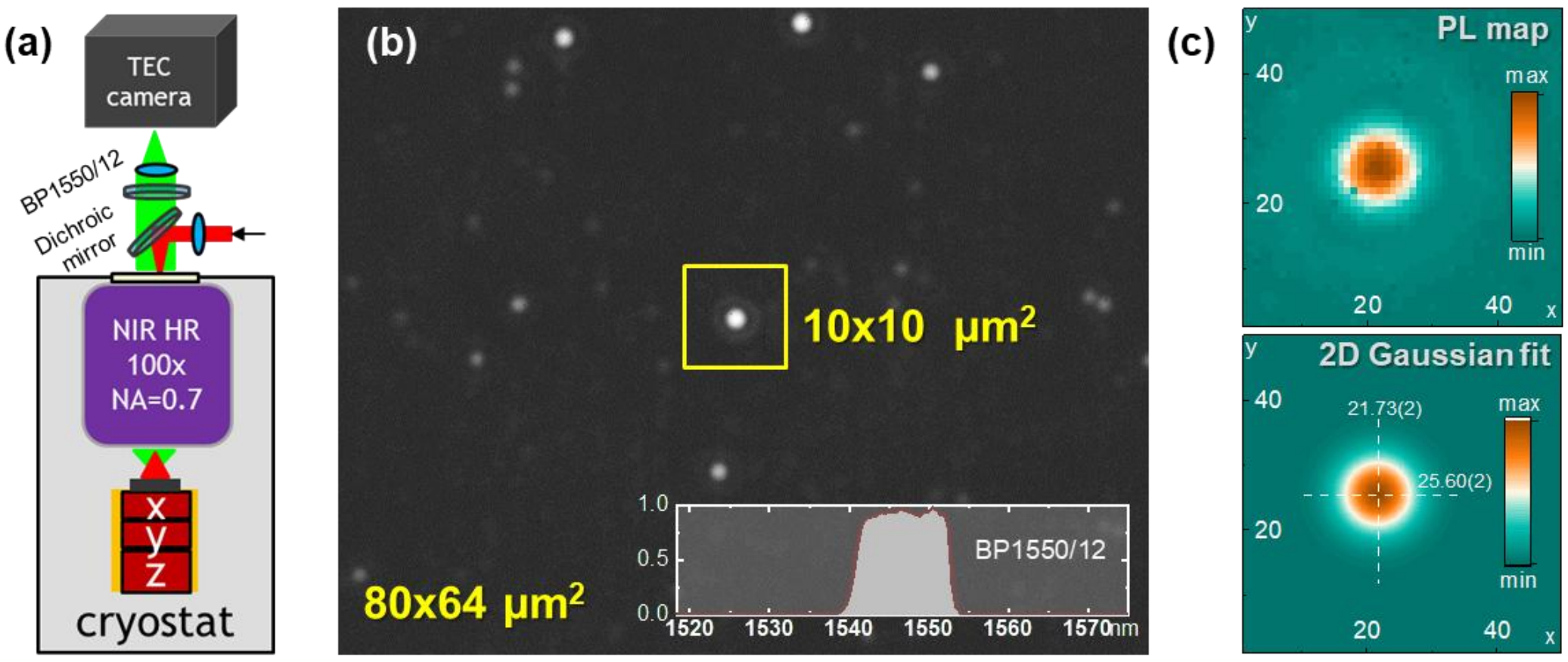

Figure 3 (a) Schematic of wide-field photoluminescence (WFPL) setup. (b) WFPL image after continuous-wave laser illumination at 640 nm. The expanded laser beam allows uniform excitation over the entire objective field-of-view. Bright spots correspond to QD emission enhanced by the DBR, matched to the 1550/12 nm bandpass filter in front of the 2D InGaAs thermoelectrically cooled monochrome camera. (c) Enlarged image of the preselected QD. The fitting with a 2D Gaussian function reveals a centroid accuracy of ±0.02 px (±2.5 nm) and yielding an FWHM of 1.25 µm, which is near the diffraction limit.

**Microphotoluminescence spectroscopy and quantum optics**

The QD selected by wide-field PL imaging was first characterized by power- and polarization-resolved high-spatial-resolution photoluminescence (µPL) under continuous-wave (CW) 805 nm above-bandgap excitation (**Figure 4(a)**; see extended data in Supp. Inf., Sec. A5). At low excitation power, the spectrum contains several well-resolved spectral lines between 1540 and 1552 nm. The line near 1540 nm exhibits a fine-structure splitting (FSS) of roughly 70 µeV, attributed to a neutral exciton (X) transition. The non-zero splitting is related to the in-plane anisotropy of the confinement potential, which can arise from QD geometry, intermixing, strain, or piezoelectricity, and is a natural consequence of the SK growth mode. A second transition at nearly 1546 nm shows similar polarization splitting, antiphase to the exciton line, with an additional superlinear power dependence, consistent with the biexciton (XX) emission. The most intense line at 1549.5 nm exhibits approximately linear power dependence and no resolvable FSS. Its spectral separation of approximately 5.5 meV from X suggests that the line is a negatively charged exciton (X-). Additional lower-energy transitions appearing at elevated excitation powers are attributed to higher-order charged complexes but are not assigned further here. We note that a similar spectral pattern was found in multiple other QDs that fall within selected spectral filtering.

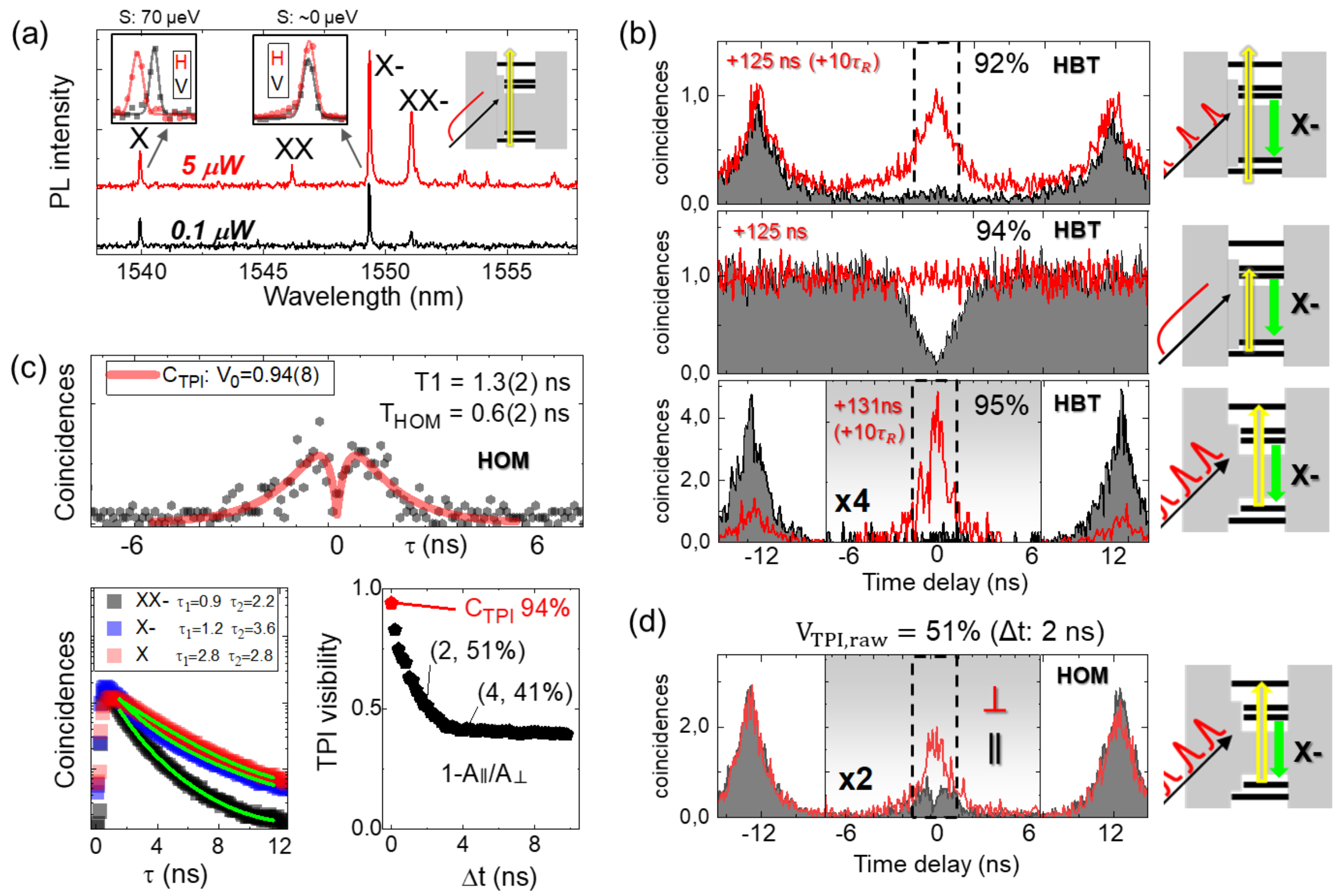

Figure 4. (a) µPL spectra observed for a single QD within a spectral filter 1550/12 nm. Inset: polarization- resolved spectra on the linear basis (H - horizontal and V - vertical) showing the X-line FSS and no FSS for the X-. (b) HBT results for the X- line under three excitation schemes: (top panel) pulsed non-resonant, (middle) CW, (bottom) pulsed quasi-resonant. (c) TPI results: fit to the experiment results obtained in (d). (d) TPI results under pulsed quasi-resonant excitation. (e) TRPL results.

The X- emission was subsequently coupled to a single-mode fiber, spectrally isolated with a tunable filter with roughly 0.3 nm resolution, and detected with two superconducting nanowire single-photon detectors. Three excitation schemes were tested to determine the conditions providing the most favorable balance between photon count rate, background suppression, and multiphoton probability (**Figure 4(b)** and Supp. Inf., Sec. A5, Fig. S5): (i) pulsed non-resonant ($f_{rep}$=80MHz, 805 nm); (ii) CW at 1536 nm; (iii) pulsed quasi-resonant ($f_{rep}$=76.34 MHz, 1460 nm). The second-order autocorrelation values ($g^{(2)}(0)$) reported here were evaluated directly from the measured coincidence data without background subtraction or detector response deconvolution.

Under the (i) excitation scheme, the system is populated with hot charge carriers excited in the InP barrier, which might result in increased spectral wandering from local electric field fluctuations, background emission n due to wetting layer recombination, and population of higher-order excitonic states, all of which might reduce the purity of the single photon emission. In this case, we observe X- count rate saturated at roughly 20 kHz, and competing lower-energy charged transitions reach approximately 10 kHz, indicating that nonresonant carrier capture distributes the population among several excitonic complexes. The corresponding pulsed autocorrelation histogram displays pronounced antibunching. However, it yields a limited raw $g^{(2)}(0)$=0.081±0.003, evaluated by integrating the central and lateral peaks over a 2 ns time window. Time-resolved photoluminescence (TRPL) measurements (**Fig. 4(c)**) reveal biexponential decay dynamics, with a fast X- component of approximately 1.2 ns, which defines the principal timescale of the emitted photon wavepacket.

We next applied the (ii) excitation scheme. The laser line was set approximately 7 meV above the X- emission energy, exciting specific QD energy levels via phonon scattering. Spectral tuning of the excitation, together with laser power control, enabled a maximum X- count rate of about 35 kHz, indicating that excess electron capture is highly possible while additional excitonic complexes near the X- state are optically active. The autocorrelation function exhibits a clear antibunching dip with a recovery time of roughly 1.5 ns. Averaging coincidences within a 0.5 ns interval around zero delay yields a raw $g^{(2)}(0) = 0.064\pm0.002$. Because this value was obtained under CW excitation and with a different integration interval, it cannot be directly compared quantitatively with the pulsed measurements.

For the (iii) excitation scheme, the energy detuning of approximately 49 meV from the X- is consistent with excitation via a higher QD state manifold. Moreover, the laser energy omits wetting layer states as well as the InGaAsP layers composing the DBR. The X- count rate saturates near 18 kHz, and the spectrum shows reduced background from neighboring spectral lines (Supp. Inf., Sec. A5, Fig. S5c). For the quantum statistics measurements, the excitation power was reduced to roughly 12 kHz to further maximize the signal-to-noise ratio. The resulting autocorrelation histogram yields a raw $g^{(2)}(0) = A_{0,2ns}/A_{10,2ns} = 0.059 \pm 0.002$, using identical 2 ns integration windows ($\tau = 0$ ns => $A_{0,2ns} = \sum_{-1}^{+1} b_i$; $b_i$ – bin amplitude; $\tau_r = \frac{10}{frep}$ ≅131ns => $A_{10,2ns} \sum_{\tau_r-1}^{\tau_r+1} b_i$; $u_{g2(0)} = g^{(2)}(0)(1/A_{0,2ns}+1/A_{10,2ns})^{1/2} = 0.002$). The normalization performed for correlation peak at delay $\tau$ greater than 131 ns (×10 laser repetition rate) was used to ensure no influence of blinking-induced increase of adjacent peaks – the blinking timescale is around 39 ns.

Two-photon interference (TPI) measurements were conducted in the Hong-Ou-Mandel (HOM) configuration, under the (iii) excitation scheme, with fiber-based electronic polarization control on the delayed arm of the Mach-Zehnder interferometer. The delay was adjusted to compensate for the excitation laser's nominal repetition rate $f_{rep}$. **Figure 4(d)** presents the raw coincidence histograms for the two polarization settings. A pronounced suppression of the central coincidence feature is observed for parallel polarizations, indicating TPI from the X- emission at 1549.5 nm. The HOM correlation histogram was fitted with a coherence-based model ($C_{TPI}(\tau) = A(\exp(-|\tau|/T_1) - V_0 \exp(-2|\tau|/T_{HOM}))$), yielding $T_1 = 1.3\pm0.2$ ns, the HOM decay parameter, $T_{HOM} = 0.6\pm0.2$ ns, and zero-delay visibility $V_0 = 0.94\pm0.08$ (**Figure 4(c)**). The obtained $T_1$ time is consistent, within an uncertainty of ±0.2 ns, with the direct result from decay time measurement.

A more extensive analysis of the HOM result requires evaluation of the interference visibility across the entire photon wave packet, rather than post-selected interference at $\tau = 0$ , which is unaffected by dephasing and limited by the excited state lifetime. The raw integrated TPI visibility as a function of time window has been obtained using the formula

$$V_{HOM} = 1 - \frac{A_{\parallel}(\Delta t)}{A_{\perp}(\Delta t)}$$

and plotted in **Figure 4(c)**, where $A_{\parallel}$ and $A_{\perp}$are the integrated raw coincidence areas within the same time window $\Delta t$ for parallel and orthogonal polarizations. This quantitative analysis shows approximately 51% TPI visibility for a 2 ns time range, decreasing to 41% for time ranges longer than 4 ns.

For comparison, an MBE-grown InAs/InAlGaAs quantum dot (QD) integrated with a circular Bragg grating exhibited a raw two-photon-interference visibility of 71.9% within a 3 ns window [47] and record high value of 91.7% for deterministic measurement scheme requiring modulation of the polarized photons [31]. A closely related

MOVPE-grown InAs/InP QD incorporated into a mesa with a bottom metallic reflector achieved 35% within a 4 ns window [46], while a planar MOVPE-grown InGaAs QD on an AlAs/GaAs distributed Bragg reflector (DBR) showed a visibility of 14% [43]. Although differences in excitation conditions, integration windows, and photonic environments preclude a strictly quantitative comparison, the approximately 51% visibility obtained here within 2 ns indicates that the unprocessed planar InAs/InP–InGaAsP/InP platform already supports competitive photon coherence without intentional Purcell enhancement. Deterministic integration into a resonant structure could further increase indistinguishability through radiative-rate enhancement, while controlled doping or electrical stabilization may suppress charge-noise-induced dephasing. The relevant systems are summarized in **Table 1**

| QD Type | Growth | Device | Excitation | Purcell factor | Excitonic state | $V$HOM | $T$1 | ΔE FWHM | Refs. |
|---|---|---|---|---|---|---|---|---|---|
| InAs/InP | MOVPE | Planar/DBR | QR (p-shell) | 1 | X- | $V_{2ns}$=51(1)% $V_{4ns}$=41(1)% | 1300 ps | 44.2 μeV | This work |
| InAs/InAlGaAs | MBE | CBG/Au | LA | 4.7(5) | X+/- | $V_{3ns}$=91.7(2)% | 257.5 ps | 28.9 μeV | [31] |
| InAs/InAlGaAs | MBE | CBG/Au | QR (p-shell) | 4.7(5) | X+/- | $V_{3ns}$=71.9(2)% | 257.5 ps | 28.9 μeV | [47] |
| InAs/InP | MOVPE | Mesa/Al | TPE | 1 | XX | $V_{4ns}$=35% | 340 ps | 47 μeV | [46] |
| InAs/MMB | MOVPE | CBG/Au | LA, SUPER | 4.3(8) | X+ | $V_{fit}$=34.9%(LA) $V_{fit}$=10.4%(SUPER) | 500 ps | 20.0 μeV | [50] |
| InAs/InP | MOVPE | CBG/Al | LO | 5 | X+/- | $V_{fit}$=19.3% | 530 ps | 87 μeV | [43] |
| InAs/MMB | MOVPE | CBG/Au | p-shell | 3.0(7) | X+ | $V_{fit}$=8.1% | 520 ps | 65 μeV | [51] |
| InAs/MMB | MOVPE | Planar/DBR | RF | 1 | X+ | $V_{fit}$=14.4% | 1010 ps | ≈30 μeV | [52] |

Table 1. Reported two-photon interference visibility metrics (VHOM) for a single QD emitting in the telecom C-band, with experimental conditions: QD material system, QD growth method (MOVPE: metal-organic vapor phase epitaxy, MBE: molecular beam epitaxy), device type (DBR: distributed Bragg reflector, MMB: metamorphic buffer, CBG: circular Bragg grating), excitation scheme (QR: quasi-resonant excitation, TPE: two-photon excitation, LA: LA-phonon assisted excitation, SUPER: swing-up population quantum emitter population, LO: LO-phonon assisted excitation, RF: resonant fluorescence), state lifetime (T1), linewidth (ΔE). Subscript of the visibility indicates the integration window size, while “fit” states for analysis using full peak amplitude.

**Summary**

In conclusion, a monolithic telecom C-band quantum-emitter platform has been demonstrated, utilizing low-density InAs/InP quantum dots (QDs) grown on a 20.5-pair InGaAsP/InP distributed Bragg reflector. The DBR was designed with experimentally calibrated structural and optical parameters to achieve broadband upward photon extraction while preserving the native InP wafer and keeping the radiative Purcell factor close to unity. The calculated upward extraction efficiency is approximately 13%, four times higher than that of an equivalent emitter in bulk-like InP, and corresponds to the spectral range of the low-density QD ensemble.

The enhanced photon flux from the unprocessed planar structure enables wide-field preselection of individual C-band emitters. For the quantum dot under investigation, fitting the diffraction-limited photoluminescence (PL) image yields a conservatively estimated statistical centroid-localization precision of approximately 2.5 nm. While this value reflects optical fitting precision rather than full QD-to-device registration accuracy, it demonstrates that the mirror-assisted signal is sufficient to initiate a marker-based deterministic fabrication workflow.

Spectroscopic characterization reveals a charged-exciton transition at 1549.5 nm. Among the excitation schemes evaluated, pulsed quasiresonant excitation offers the most favorable balance between count rate and multiphoton suppression, resulting in a raw $g^{(2)}(0)$ value of 0.059(2). Consecutively emitted photons display a raw Hong–Ou–Mandel visibility of approximately 51% within a 2 ns integration window and approximately 40% over the entire detected wavepacket. To the best of current knowledge, the latter value ranks among the highest reported full-wavepacket visibilities for pulsed InAs/InP C-band emission without intentional resonant Purcell enhancement.

The significance of this platform resides in its integration of three sequential requirements within a single epitaxial structure: sufficient broadband extraction for emitter identification, preservation of high photon quality prior to processing, and compatibility with subsequent monolithic nanofabrication. The DBR enhances upward photon collection during emitter preselection while remaining an integral part of the wafer for subsequent deterministic nanofabrication. Future marker-based fabrication of aligned nanocavities, microlenses, mesas, vertical cavities, or related structures might further improve mode selectivity, extraction efficiency, and radiative rate, while making use of the high pre-fabrication coherence demonstrated in this work.

## ACKNOWLEDGEMENTS

This work was supported by the National Science Centre, Poland, under Projects No. 2023/50/E/ST7/00581 and 2025/57/B/ST7/04916 (TSUNAMI). P.H. acknowledges support from the Polish National Agency for Academic Exchange under the Polish Returns programme. The authors thank Paweł Wyborski for advice on polarization control in the two-photon-interference measurements, Elizaveta Semenova for support with epitaxial growth, and Jarosław Serafińczuk for X-ray diffraction measurements.